\documentclass[conference]{IEEEtran}
\usepackage{graphicx} % Required for inserting images
\usepackage{amsmath}
\usepackage{cleveref}
\usepackage{xcolor}
\usepackage{framed}
\usepackage{ascii}
\usepackage[T1]{fontenc}
\usepackage[scaled]{beramono}
\usepackage{listingsutf8}

\title{LLM-Aided Compilation for Tensor Accelerators}

\author{
\IEEEauthorblockN{Charles Hong\IEEEauthorrefmark{1}, Sahil Bhatia, Altan Haan, Shengjun Kris Dong, Dima Nikiforov, Alvin Cheung, Yakun Sophia Shao}
\IEEEauthorblockA{
\textit{University of California, Berkeley}\\
Berkeley, CA, USA\\
\IEEEauthorrefmark{1}charleshong@berkeley.edu\\
}}
\date{March 2024}

\begin{document}

\IEEEoverridecommandlockouts
\IEEEpubid{\makebox[\columnwidth]{ 979-8-3503-7608-1/24\$31.00 \copyright2024 IEEE \hfill} \hspace{\columnsep}\makebox[\columnwidth]{ }}

\maketitle

\begin{abstract}
Hardware accelerators, in particular accelerators for tensor processing, have many potential application domains. However, they currently lack the software infrastructure to support the majority of domains outside of deep learning. Furthermore, a compiler that can easily be updated to reflect changes at both application and hardware levels would enable more agile development and design space exploration of accelerators, allowing hardware designers to realize closer-to-optimal performance. In this work, we discuss how large language models (LLMs) could be leveraged to build such a compiler. Specifically, we demonstrate the ability of GPT-4 to achieve high pass rates in translating code to the Gemmini accelerator, and prototype a technique for decomposing translation into smaller, more LLM-friendly steps. Additionally, we propose a 2-phase workflow for utilizing LLMs to generate hardware-optimized code.
\end{abstract}
\section{Introduction}
Hardware accelerators~\cite{nvidia_cuda, tpu-isca2016, aws-inferentia} have become a critical driving force for the recent breakthroughs~\cite{alexnet, goodfellow2014generative, resnet, transformer, chatgpt} in artificial intelligence and machine learning. They provide hundred-fold improvements in performance and energy efficiency in running deep neural networks (DNNs). 
With the proliferation of new TA designs, the number of compilers and domain-specific languages (DSLs) has also exploded.
For deep learning applications, compilers like XLA and TVM provide end-to-end support for the popular deep learning frameworks PyTorch, JAX, and TensorFlow frameworks which are used to implement most DNN software~\cite{xla, chen2018tvm}.

However, these accelerators are not only useful for processing DNNs. For example, the systolic array architecture at the heart of many of these accelerators has long been known to be useful for a wide range of tensor-related computations, such as tensor decomposition~\cite{kung1979systolic}. 
Furthermore, recent work suggests that these accelerators, which we call tensor accelerators (TAs), have promise in accelerating a range of applications ranging from graph algorithms like PageRank to partial differential equations for financial modeling \cite{hsu2021accelerating, zhang2022simd2}. 

As shown by these works, in order to leverage the performance benefits of TAs, applications must be compiled to primitive operations in the domain-specific language (DSL) supported by one specific TA, and in order for this to occur the TA must first support the key operators of the application. 
The development of both applications and accelerators is limited by this cyclical dependency. 
Adapting existing code to DSLs requires developers to manually translate the code or even build custom compilers, which must be modified each time the hardware backend changes.

An ideal compiler framework can adapt to changes both above it (application-level) and below it (architecture- or microarchitecture-level). Recent work demonstrates the impressive performance of large language models (LLMs) in various code-related tasks~\cite{chen2021humaneval, nijkamp2022codegen, jain2024livecodebench, patil2023gorilla}, as well as general reasoning ability and instruction-following~\cite{brown2020fewshot,openai2024gpt4}. However, it is unclear how LLMs perform in code analysis and generation for DSLs with little to no presence in their training corpora. In this work, we investigate how LLMs can be used in an agile compiler framework for hardware accelerators, and propose that optimizing compilers could be implemented in a two-phase flow.
The first phase involves translating the given source program to a functionally correct implementation in the DSL, ensuring functional correctness. The second phase focuses on optimizing the DSL code using a cost model-driven search approach to maximize performance on the target hardware accelerator.

\section{Related Work}
% \todo{Sahil add things related to code gen}
Code translation is essential for keeping software workflows updated with recent DSLs and optimizations. Existing approaches include pattern matching-based compilers~\cite{mold}, search-based techniques~\cite{metalift}, and neural methods~\cite{ngst}. However, these approaches require significant human effort to develop and maintain, and often struggle to scale to complex domains. In this work, we leverage LLMs success in code generation~\cite{codellama} and optimization~\cite{cummins2023compiler} to explore their potential for generating optimized code for TAs. 

% \todo{accelerator optimizations like DOSA}
There has been a significant amount of work in exploring the performance spaces of hardware and software implementations for TAs. However, while significant performance improvements are possible, such design space exploration techniques are often limited to finding an optimal point within a given search space~\cite{vaesa, dosa}, or use abstractions from which bridging the gap to real systems is difficult~\cite{parashar2019timeloop}. Automatic compilers robust to application and hardware changes will allow designers to quickly modify search spaces without significant compiler update efforts.

% \todo{MLIR and other abstractions}
% BLAS, Intel AMX, ARM SME, other academic works ($\text{SIMD}^2$). \todo{expand on these examples}
% This work differs from works that propose abstractions for tensor computation because \todo{...}
Existing systems for tensor computation make use of abstractions like the BLAS library or, as in the case of Halide, use DSLs to represent computations in a portable and scheduling-friendly manner~\cite{jrk2013halide}. In this work, we focus on enabling TA developers to compile code to accelerators as quickly and easily as possible, so we elide the addition of heavy infrastructure that would add burden to the developer workflow. However, we are not opposed to the use of intermediate representations or other abstractions when building LLM-aided compilers, and believe that development in such a direction will enable LLM-aided compilation and optimization to be carried out in a more systematic and verifiable manner.

\section{Proposed Methodology}
\begin{figure*}[!ht]
\vspace{-0.02in}
\centering
\includegraphics[scale=0.5,trim=0cm 5cm 0cm 4.15cm, clip]{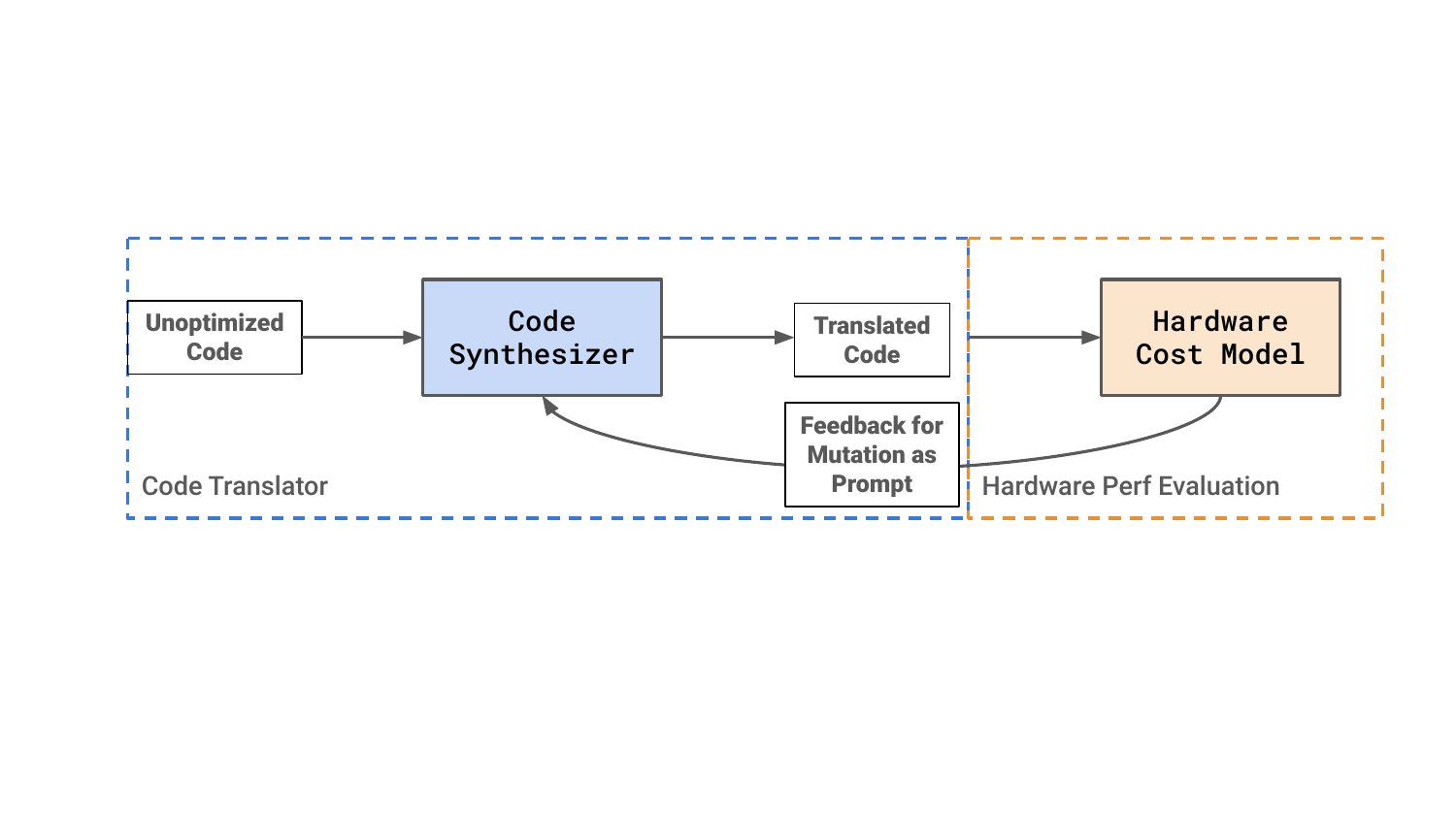}
\vspace{-0.1in}
\caption{An overview of our proposed framework. 
% The framework consists of two components, a search-based technique for translating unoptimized code to the APIs of the tensor accelerator and a cost-model for evaluating the performance of the translated code and returning the feedback to the search component. The translator would incorporate the feedback from the cost-model and return the updated version of the code. 
}
\label{fig:framework}
\vspace{-0.1in}
\end{figure*}

\subsection{Overview}
In this section we provide an overview of our two phase approach. In \Cref{fig:framework} shows our proposed workflow which integrates the search-based translation and hardware cost models. Our workflow involves a code synthesizer generating potential translations for a general-purpose code, which are then verified for functional equivalence with the source program using test-cases. Subsequently, the verified code is passed to the cost model, which offers concrete feedback indicating changes the synthesizer should apply to the generated code. 
% A potential advantage of using LLM is their text-based interface which serves as a unified means of providing multimodal feedback, such as performance numbers and specific program components that could be mutated, as natural language. This is particularly important due to the nature of hardware modeling---this cost model could be a fast but lower-fidelity analytical model, or a slow but cycle-accurate RTL simulation. It might provide only a cycle count, or, as we shown in prior work such as DOSA~\cite{dosa}, might be a differentiable model that provides a gradient pointing in the direction of performance improvement. An iterative loop between code generation, verification, evaluation and feedback ensures continuous refinement of the translated code, optimizing not only for functionality but also for performance.
% on specific accelerators. 

\subsection{Code Template Generation with LLMs}
\label{sec:translation_proposed}
% \todo{Charles: Sahil can you explain how you know LLMs are better for generating code templates than solver-based implementations, and why this might be the case?}

There are two general approaches to building code translators: symbolic and neural. Symbolic approaches include building pattern-matching compilers, for which rules can be painstaking to manually specify and maintain. To address this, verified lifting~\cite{metalift} uses search followed by verification to find a functionally equivalent implementation of the source program in the target language. However, most lifting-based approaches rely on symbolic solvers that use strategies like enumerative or constraint-based search to perform the translation. Scaling symbolic search requires significant effort and domain knowledge, as users must explore heuristics such as type-based filtering, template enumeration, and multi-phase synthesis to shrink the search space. 

LLMs have emerged as a promising alternative to symbolic approaches.
% LLMs have shown impressive performance in programming-related tasks such as code generation, summarization, and translation, among others. 
These models have been trained on massive amounts of code data from sources such as documentation and code repositories, which potentially allows them to learn about the syntax and semantics of various programming languages. 
% One of the key strengths of LLMs is their ability to excel at reasoning about new tasks without the need for additional training, leveraging the context provided in the prompt to adapt to various scenarios as demonstrated in recent work~\cite{brown2020fewshot}. 
% For instance, recent work~\cite{few-shot} demonstrated that LLMs are capable of few-shot reasoning, which highlights their adaptability to new tasks with minimal examples.
TAs and other DSAs present a unique challenge because their low-level programming languages have little to no presence in LLMs' training corpora.
We propose that LLMs can nonetheless be leveraged to simplify the process of generating optimized code for TAs by exploiting their contextual reasoning capabilities~\cite{brown2020fewshot} and decomposing the problem into multiple semi-structured steps.
% we aim to develop an efficient and scalable code translation approach that can generate code tailored to specific hardware architectures.

To guide the LLM in generating the desired target code, we provide a structured prompt that
% (shown in ~\cref{fig:prompt}) 
% that includes the necessary information for the translation task. The prompt 
consists of three main components: instructions, target language specification, and the source program. The instruction section contains a high-level description of the task, specifying the goal of translating the source program to the target language. The target language specification section enumerates the available operators and constructs in the target language, and optionally provides example programs in the target language, providing the LLM with the necessary context about the TA. Finally, the source program section includes the high-level code that needs to be translated. Appendix~\ref{sec:prompts} shows an instantiation of this prompt structure demonstrating how the components are populated with specific details, and \Cref{sec:experiments_gemmini} discusses how well these prompts work.
% for translation to the Gemmini ISA. 
% Our approach seeks to overcome the limitations of symbolic search methods by leveraging the generalization abilities of LLMs, potentially reducing the need for extensive domain knowledge and heuristics in the code translation process.

% \begin{figure}[]
% \centering
% \includegraphics[scale=0.45,trim=0cm 0cm 0cm 0cm, clip]{figs/prompt-struc-hardware.png}
% \caption{High-level prompt structure used for code translation.}
% \label{fig:prompt}
% \vspace{-0.1in}
% \end{figure}

\subsection{Cost Model-Driven Code Translation}
In addition to being correct, compiler-generated code should be performant, especially when the target is hardware accelerators meant to improve application latency and efficiency.
% Evaluating program performance on TAs before tape-out is extremely computationally expensive, as RTL designs must be simulated signal-by-signal.
For performance optimization, we propose a search-based technique for code translation that integrates feedback from a TA cost model. Previous work, such as Ansor~\cite{ansor}, has successfully implemented cost model-based approaches for scheduling tensor operations in the TVM compiler framework. While these approaches are effective, they rely on manually designed search spaces and require extensive training for each new hardware target. In contrast, we seek to leverage LLMs' knowledge about programs to optimize a less structured space. 
% By utilizing LLMs, we aim to reduce the need for manual search space design and enable more efficient adaptation to new hardware targets.

In particular, we suggest an iterative and hierarchical approach. First, we prompt the model with a set of possible optimizations, such as combining instructions or changing data movement patterns. The model is then asked to optimize each block of computation based on the available optimizations and any latent knowledge about program optimization. We experiment with two approaches: 1) having the LLM directly generate the optimized code, and 2) having the LLM generate Halide-style scheduling operations~\cite{jrk2013halide}.
Next, we prompt the model to generate the optimal ordering of these blocks in the final program. If the LLM understands dependencies and program performance, it can propose an efficient arrangement of the blocks, taking into account factors such as data locality and parallelism. Program performance (and potentially other indicators, like hardware counters), generated from a cost model or by running the code, can be used as feedback, providing domain-specific information for the model to iteratively refine its optimization decisions.

\section{Experiments}
In this section, we discuss a number of experiments that explore the feasibility of utilizing LLMs for various parts of the accelerator compilation flow. Based on these experiments, we discuss the most effective strategies we observed and suggest directions for future exploration. 
% However, our approach is general and we plan to extend it to other hardware accelerators in future work.

\subsection{Translating Robotics Kernels to Gemmini's ISA}
\label{sec:experiments_gemmini}

Robotics is a driving application with rising interest, both across the scientific community and in relation to hardware acceleration. Prior work has investigated the potential for custom hardware accelerators to speed up key robotics kernels~\cite{neuman2019benchmarking, nikiforov2023rose}. These kernels are a target for acceleration via systolic array accelerators on the edge, due to their latency sensitivity and heavy use of matrix operations.
However, implementing performant code for such applications on custom hardware is difficult due to the lack of compilers, and building general-purpose libraries can actually result in control flow-heavy code with worse performance than an assembly implementation.
In this section we demonstrate translation of general-purpose matrix code from these kernels to the instruction-set architecture (ISA) for Gemmini~\cite{gemmini}, an academic systolic array accelerator. The prompt describing the set of functions available to the LLM, which effectively represent Gemmini's ISA, is given in Appendix~\ref{sec:prompts}.

We use a simple test to determine whether a correct result has been generated. However, these results do not provide the model with access to the test cases, nor is there yet a feedback loop of test results to code generation. So, it is highly unlikely that model outputs are overfit to our current set of test cases.

\subsubsection{Model-Predictive Control (Matrix-Vector Operations)}
\label{sec:matvec}

We begin by translating a kernel from the TinyMPC model-predictive control implementation~\cite{nguyen2024tinympc}. Specifically, we focus on the backward pass, which contains four matrix-vector multiplications of different sizes. The sizes reflect those for a quadrotor drone; the largest operation comprises a 12$\times$12 matrix multiplied by a 12$\times$1 vector. We use a configuration of Gemmini with a 4$\times$4 systolic array, meaning that such a computation requires at least 9 compute instructions, plus instructions to configure the accelerator and move data into and out of local memory.
For this kernel, we translate one matrix-vector multiplication at a time, following the strategy from Section~\ref{sec:translation_proposed}. 

We ablate a number of prompting techniques. This is to 1) evaluate the importance of each component of our prompt, and 2) evaluate the effectiveness of the LLM's (specifically, \texttt{gpt-4-turbo}'s instruction following and in-context learning (ICL)~\cite{brown2020fewshot} in the context of accelerator DSLs with little information available in pre-training data.

Specifically, we explore the following options:
\begin{itemize}
    \item \textbf{Zero-shot}: We try generating Gemmini code with only ISA descriptions and no implementation example, to evaluate the LLM's ability to generate code based solely on pre-training and its ability to reason about the functions in the specification.
    \item \textbf{One-shot}: We provide a single Gemmini code example for a matrix-vector multiplication. This boosts code correctness significantly, and qualitatively reduces the variance in generated code hugely. Because generated code follows the style of the provided examples, syntax and other compilation errors also decrease significantly. Note that for all cases, we evaluate pass rate on problems other than the one used for this example.
    \item \textbf{NL annotation}: We annotate the one-shot example with inline natural language (NL) comments explaining each function call and its arguments. We find that this technique improves the ability of the LLM to reason about the provided functions, and extrapolate implementations that are different from the provided example while following its style.
    \item \textbf{No ISA}: We remove the ISA (target language) specification from the prompt. The results establish that it is an essential part of the translation flow.
    % \item \textbf{Choice of LLM}: We evaluate both \texttt{gpt-4-turbo} (Table~\ref{tab:matvec_nl_input}) and \texttt{gpt-3.5-turbo} (Table~\ref{tab:matvec_3.5}) as of April 9, 2024, which illustrates the improvement in reasoning for GPT-4 compared to older models.
\end{itemize}

\begin{table}[htbp]
\centering
\begin{tabular}{p{0.45\linewidth}|ccc}
\hline
\multicolumn{1}{c}{} & \multicolumn{3}{c}{pass@k} \\
\multicolumn{1}{c}{\textbf{}} & k=1 & k=10 & k=50 \\ \hline
\textbf{Zero-shot} & 0.33\% & 3.33\% & 16.7\% \\ \hline
\textbf{One-shot ICL} & 44.67\% & 84.42\% & 99.79\% \\ \hline
\textbf{One-shot ICL (NL-annotated)} & 46.0\% & 88.81\% & 99.98\% \\ \hline
\textbf{No ISA, One-shot ICL (NL-annotated)} & 1\% & 9.12\% & 29.29\%  \\ \hline
\end{tabular}
\vspace{0.1in}
\caption{\texttt{gpt-4-turbo} translated code correctness for matrix-vector multiplications, with natural language descriptions for functions in the input code.}
\vspace{-0.1in}
\label{tab:matvec_nl_input}
\end{table}

% \begin{table}[htbp]
% \centering
% \begin{tabular}{p{0.45\linewidth}|ccc}
% \hline
% \multicolumn{1}{c}{} & \multicolumn{3}{c}{pass@k} \\
% \multicolumn{1}{c}{\textbf{}} & k=1 & k=10 & k=50 \\ \hline
% \textbf{Zero-shot} & 0\% & 0\% & 0\% \\ \hline
% \textbf{One-shot ICL (NL-annotated)} & 32.0\% & 54.06\% & 94.0\% \\ \hline
% \end{tabular}
% \vspace{0.1in}
% \caption{\texttt{gpt-3.5-turbo} translated code correctness for matrix-vector multiplications, with natural language descriptions for functions in the input code.}
% \vspace{-0.1in}
% \label{tab:matvec_3.5}
% \end{table}

% \begin{table}[htbp]
% \centering
% \begin{tabular}{p{0.45\linewidth}|ccc}
% \hline
% \multicolumn{1}{c}{} & \multicolumn{3}{c}{pass@k} \\
% \multicolumn{1}{c}{\textbf{}} & k=1 & k=10 & k=50 \\ \hline
% \textbf{One-shot} & 0\% & 0\% & 0\% \\ \hline
% \textbf{LLM-based hole-filling} & 32.0\% & 54.06\% & 94.0\% \\ \hline
% \end{tabular}
% \vspace{0.1in}
% \caption{\texttt{gpt-4-turbo} translated code correctness for problem 8, which could not be translated in one shot.}
% \vspace{-0.1in}
% \label{tab:prob8}
% \end{table}

We additionally explore whether it is more useful to provide NL descriptions, or full implementations of functions in the input (general-purpose) code. We implement matrix-vector multiplication as the more general matrix-matrix multiplication. Even though this is a very common operation, translation correctness improves in both zero-shot and one-shot scenarios with code implementations. This is consistent with previous results, as we suspect that like our one-shot example, a general-purpose implementation provides structure for the LLM to follow in its response. 

As shown in Table~\ref{tab:matvec_source_type}, providing both NL and code hurts correctness of generated code in both zero-shot and one-shot cases, showing that increasing prompt size without providing new information may degrade code generation performance.

\begin{table}[htbp]
\centering
\begin{tabular}{p{0.45\linewidth}|ccc}
\hline
\multicolumn{1}{c}{} & \multicolumn{3}{c}{pass@k} \\
\multicolumn{1}{c}{\textbf{}} & k=1 & k=10 & k=50 \\ \hline
\textbf{NL only} & 46.0\% & 88.81\% & 99.98\% \\ \hline
\textbf{Semantics only} & 50.67\% & 92.23\% & 100\% \\ \hline
\textbf{Semantics and NL} & 46.33\% & 87.48\% & 99.96\% \\ \hline
\end{tabular}
\vspace{0.1in}
\caption{\texttt{gpt-4-turbo} translated code correctness for matrix-vector multiplications, with different presentations of source code functions.}
\vspace{-0.1in}
\label{tab:matvec_source_type}
\end{table}

\subsubsection{Riccati Recursion (Matrix-Matrix Operations)}
\label{sec:matmul}

Next we translate C++ code for Riccati recursion, a well-known method for solving the finite-horizon discrete time linear quadratic regulator (LQR) problem~\cite{tedrake2023underactuated}. Specifically, we focus on implementing seven matrix-matrix multiplications of various sizes and types, one of which is used for our prompting example. In some cases, matrices are multiplied; in some cases, a bias matrix is added or subtracted from the result. The largest is a 36$\times$36 and 36$\times$12 matrix-matrix multiplication, with state and action space sizes based on a quadrotor drone (assuming an action space size of 4) and a quadruped~\cite{neuman2019benchmarking}.

Due to cost constraints, we replicate only the best-performing of experiment from Section~\ref{sec:matvec}, that with reference implementations for input code, as well as an NL-annotated in-context example. We next compare the case of providing a matrix-vector example with the case of providing two matrix-matrix examples, one with a transposed matrix and a bias, and one without. Providing both examples does not boost pass rate, but we note that the LLM performs better when examples are provided \emph{before} other instructions in the prompt. Table~\ref{tab:matmat} shows \texttt{gpt-4-turbo}'s pass rate for these 6 functions. Ultimately, we are able to generate correct code for 5 out of 6 test functions.
% the number of solved functions actually decreases to 4, perhaps because the generated result is overfit to the example.

\begin{table}[htbp]
\centering
\begin{tabular}{p{0.45\linewidth}|ccc}
\hline
\multicolumn{1}{c}{} & \multicolumn{3}{c}{pass@k} \\
\multicolumn{1}{c}{\textbf{}} & k=1 & k=10 & k=50 \\ \hline
\textbf{One-shot ICL (Matrix-vector example, NL-annotated)} & 2.33\% & 20.55\% & 64.51\% \\ \hline
\textbf{One-shot ICL (Matrix-matrix example, NL-annotated)} & 1.67\% & 15.21\% & 51.21\% \\ \hline
\textbf{One-shot ICL (Matrix-matrix example w/ transpose and bias, NL-annotated)} & 50.05\% & 72.89\% & 89.65\% \\ \hline
\textbf{Two-shot ICL (Both examples after instructions)} & 15.55\% & 57.39\% & 81.26\% \\ \hline
\textbf{Two-shot ICL (Both examples before instructions)} & 32.41\% & 75.68\% & 83.29\% \\ \hline
\end{tabular}
\vspace{0.1in}
\caption{\texttt{gpt-4-turbo} translated code correctness for matrix-matrix multiplications.}
\vspace{-0.13in}
\label{tab:matmat}
\end{table}

\subsection{Repairing Translated Code}
\label{sec:repair}
In the previous section, we successfully translated 8 out of 9 total kernels with the prompting techniques listed. However, there was one kernel which could not be translated - this kernel multiplies a 12x4 matrix, transposed, with 4x12 matrix, not transposed, and subtracts from the result a 12x12 bias matrix. Manual inspection of generated candidates demonstrated that there were several cases of near-correct translations, where errors often occur due to incorrect addressing, strides, or constants in instruction arguments. This corresponds with prior observations that even the most powerful LLMs can struggle with arithmetic-related tasks~\cite{bubeck2023sparks}.

Prior work has addressed such challenges by fixing LLM errors such as syntactic search~\cite{lee2024guess}. For this case, we were able to produce a working result by breaking down error correct into multiple steps, i.e. by taking a close-but-incorrect candidate, prompting the LLM to locate areas of uncertainty and replace such holes with its own variable names, then produce candidates for programs with these holes filled using a set of possible constants. In this case study, this sequence of prompts was able to produce a correct result, unlike a simple prompt asking the LLM to fix the constants in its response. The specific prompts used can be found in Appendix~\ref{sec:prompts}.

\subsection{Code Optimization}

% Without fine-tuning or domain-specific pre-training~\cite{cummins2023compiler}, prompting an LLM to directly rewrite DSL code for improved performance is currently unlikely to succeed, especially if the DSL does not make up a large part of the LLM's training data. However, as most performance optimizations are generic across hardware platforms, we believe that a structured approach can enable LLMs to aid in optimization.

In this section, we describe our preliminary results with different strategies for optimizing the translated code.

\subsubsection{Structured LLM-Driven Code Rewrites}
We evaluate our hierarchical optimization process by optimizing the translated matrix-vector multiplication code described in \cref{sec:matvec}. First, we prompt ~\Cref{fig:prompt_optimize} the LLM to generate optimized versions for each block of computation in the translated code. Next, we prompt~\Cref{fig:prompt_reorder} the LLM to reorder the optimized blocks generated in the previous phase. Upon comparing the LLM-optimized code with hand-tuned code, we observed that the generated code was similar in structure and performance. LLM was able to correctly identify the optimal ordering of the blocks resulting in minimizing the data movement.

\subsubsection{LLM-driven Autoscheduling}
Instead of directly generating optimized code, we are also experimenting with using LLMs as code schedulers.
In this approach, the LLM selects schedule operations (loop reordering, etc.) to apply to the code.
%In this approach, the unoptimized code is fed into the LLM together with a set of scheduling operations (loop reordering, tiling, etc.).
% The LLM is then instructed to generate an operation to apply to the code.
% This loop is repeated to generate a full sequence of scheduling operations.
In our preliminary experiments, we use Exo \cite{exo} as the scheduling library.
Every successful Exo rewrite is guaranteed to be semantics preserving, thus eliminating correctness issues due to hallucination.
Appendix \ref{sec:llm_autoscheduling} shows \texttt{gpt-4-turbo} scheduling the \texttt{doitgen} kernel from PolyBench \cite{polybench}, a multiresolution analysis kernel from MADNESS \cite{madness}, on x86.
\section{Conclusion: Towards Agile, Automated Hardware and Compiler Co-Design}
% Enabling generation of performant domain-specific programs will open up critical research directions. 
% For example, architects are keen to use machine learning for design space exploration (DSE) of hardware and software parameters. However, most proposed DSE techniques have not been implemented in real systems. In part, this is because they typically target mathematical representations of code that are directly suited to optimization, rather than generating real programs like in code translation. 
% Code translation can also help designers integrate new accelerators into hardware systems, for example by identifying operations in general-purpose code that can be translated into more performant domain-specific code. By automating software compilation problems that are today solved ad hoc, hardware-aware, cost model-guided code translation will serve as a key component for more agile and automated accelerator design.
In this work, we demonstrate how careful prompting and breaking down compilation problems into smaller, more LLM-friendly steps can help make accelerator code translation tractable for LLMs. Specifically, we use a combination of such techniques to fully translate a set of robotics kernels to the Gemmini accelerator's ISA.
Automated code translation will speed up the accelerator design process and reduce engineering effort, by reducing the need to build and maintain compilers early on and allowing for faster evaluations of hardware.
Furthermore, we believe that future work integrating LLMs into the existing extensive corpus of tensor code optimizations will help make it easier to apply new DSLs and optimization techniques that can improve accelerator performance.
By automating software compilation problems that are today solved ad hoc, hardware-aware, cost model-guided code translation will serve as a key component for more agile and automated accelerator design.

\section*{Acknowledgements}
We thank the ISLAD reviewers for their feedback.
This work is supported in part by the National Science Foundation through grants CCF-2238346, IIS-1955488, IIS-2027575, ARO W911NF2110339, ONR N00014-21-1-2724, and DOE award DE-SC0016260, DE-SC0021982, as well as by SLICE Lab industrial sponsors and affiliates. Any opinions, findings, and conclusions or recommendations expressed in this material are those of the author(s) and do not necessarily reflect the views of the National Science Foundation.

\clearpage
\bibliography{ref}
\bibliographystyle{IEEEtran}

\clearpage
\onecolumn

\appendices

\section{Prompts for Code Translation}
\label{sec:prompts}

\lstset{
  % language=c,
  basicstyle=\ttfamily\scriptsize,
  breaklines=true,
  numbersep=3pt,
  xleftmargin=0em,
  frame=single,
  framexleftmargin=0em,
  extendedchars=\true,
  inputencoding=utf8/latin1
}

\begin{figure*}[h]
\begin{lstlisting}[language=C++, linewidth=\textwidth, basicstyle=\ttfamily\tiny]
// defined functions

#define config_ex(dataflow, act, A_transpose, B_transpose)
// configure the state of the accelerator
// dataflow is WEIGHT_STATIONARY or OUTPUT_STATIONARY
// act is the activation function, options are NO_ACTIVATION, RELU, LAYERNORM, IGELU, SOFTMAX
// A_transpose is a boolean value that represents whether the matrix A is transposed
// B_transpose is a boolean value that represents whether the matrix B is transposed

#define config_ld(cols, id)
// configure mvin instructions
// cols = number of cols in matrix in DRAM
// id = id of mvin instruction; id = 0 for mvin, 1 for mvin2, 2 for mvin3

#define mvin(dram_addr, spad_addr, cols, rows)
// mvin from DRAM to scratchpad
// mvin, configured by config_ld(..., 0)
// requires rows must be less than or equal to DIM

#define mvin2(dram_addr, spad_addr, cols, rows)
// mvin from DRAM to scratchpad
// mvin2, configured by config_ld(..., 1)
// requires rows must be less than or equal to DIM

#define mvin3(dram_addr, spad_addr, cols, rows)
// mvin from DRAM to scratchpad
// mvin3, configured by config_ld(..., 2)
// requires rows must be less than or equal to DIM

// A = input matrix
// B = weight matrix
// C = output matrix
// assume a weight-stationary dataflow

#define preload_zeros(C_acc_addr)
// preload zeros to the systolic array and set the output address in the accumulator to C_acc_addr

#define preload(B_spad_addr, C_acc_addr, B_cols, B_rows, C_cols, C_rows)
// preload weights, B
// B must be preloaded before compute
// B must have been moved in to the scratchpad first
// B_cols must be less than or equal to DIM, B_rows must be less than or equal to DIM, C_cols must be less than or equal to DIM, C_rows must be less than or equal to DIM
// must run to change the output address to C_acc_addr 
// B_spad_addr = 0xffffffff if B already preloaded

#define compute_preloaded(A_spad_addr, bias_spad_addr, A_cols, A_rows, bias_cols, bias_rows)
// compute
// A must have been moved in to the scratchpad first
// first compute after preload, does not accumulate C
// A_cols must be less than or equal to DIM, A_rows must be less than or equal to DIM, bias_cols must be less than or equal to DIM, bias_rows must be less than or equal to DIM
// bias_spad_addr = 0xffffffff if no bias
// if there is a bias, bias_cols and bias_rows are probably equal to B_cols and B_rows from preload instruction

#define compute_accumulated(A_spad_addr, bias_spad_addr, A_cols, A_rows, bias_cols, bias_rows) 
// compute
// A must have been moved in to the scratchpad first
// accumulates to same C as previous compute 
// A_cols must be less than or equal to DIM, A_rows must be less than or equal to DIM, bias_cols must be less than or equal to DIM, bias_rows must be less than or equal to DIM
// bias_spad_addr = 0xffffffff if no bias
// if there is a bias, bias_cols and bias_rows are probably equal to B_cols and B_rows from preload instruction

#define config_st(cols)
// configure mvout instruction
// cols = number of columns of matrix in DRAM

#define mvout(dram_addr, acc_addr, cols, rows)
// mvout from accumulator to DRAM
// requires rows must be less than or equal to DIM

#define fence() asm volatile("fence") 
// fence

'''
Gemmini's private memory is "row-addressed", where each row is DIM elements wide, where DIM is the number of PEs across the width of the systolic array. These elements will be of type inputType in the scratchpad, and of type accType in the accumulator.

Every private Gemmini memory address is 32 bits long. The three most signficant bits are reserved, and have special meanings:

    Bit 31 (the MSB) is 0 if we are addressing the scratchpad, and 1 if we are addressing the accumulator.
    Bit 30 is ignored if we are addressing the scratchpad, or if we are reading from the accumulator. If, instead, we are writing to the accumulator, then bit 30 is 0 if we want to overwrite the data at that address, and 1 if we want to accumulate on top of the data already at that address.
    Bit 29 is ignored if we are addressing the scratchpad, or if we are writing to the accumulator. If, instead, we are reading from the accumulator, then bit 29 is 0 if we want to read scaled-down inputType data from the accumulator, and 1 if we want to read accType data from the accumulator.
        If bit 29 is 1 for an accumulator read address, then we do not apply activation functions or scaling to the output of the accumulator.
'''
\end{lstlisting}
\caption{Gemmini ISA specification from Section~\ref{sec:experiments_gemmini}.}
\label{fig:prompt_gen}
\end{figure*}
\begin{figure*}
\begin{lstlisting}[language=C++, linewidth=\textwidth]
Gemmini is a systolic array accelerator with a scratchpad, a DIM by DIM systolic array, an accumualator, and a backing DRAM. 
The set of available functions for the Gemmini accelerator are as follows.

<insert ISA prompt here>

Your task is to rewrite the given `test` C++ Function. You need to use only the set of provided functions and constants to achieve this. The rewritten program should be semantically equivalent to the `test` function.
Please make sure that the generated code fully computes the desired operation and that the output is correct. It is essential and important that function arguments such as rows and columns should not violate constraints such as "less than or equal to".
Recall that systolic array size is 4 by 4 (DIM equal to 4) and each element is 4 bytes.
Example 1 is a simple example which should only be used for style inspiration. Write the low level code for Example 2.
\end{lstlisting}
\caption{Code translation task description, as described in Sections~\ref{sec:translation_proposed} and ~\ref{sec:experiments_gemmini}.}
\label{fig:prompt_matvec}
\end{figure*}

\begin{figure*}
\begin{lstlisting}[language=C++, linewidth=\textwidth]
Your task is to optimize the given program. The program can be optimized by reducing the number of instructions. Instructions can be reduced by minimizing the data movement, reordering computations, and merging instructions. The rewritten program should be semantically equivalent to the original program. Do not use any loops. Systolic array size is 4x4 (DIM=4) and each element is 4bytes.
// heuristics:
1. moving data ahead of time helps
2. do not remove any compute instruction unless it can merged or replaced by another instruction
3. do not remove any preload instruction unless B_spad_addr and C_spad_addr are the same as the previous preload instruction
4. number of mvin rows <= 4

<insert ISA prompt here>

<insert unoptimized block here>

\end{lstlisting}
\caption{Task description for optimizing blocks of code.}
\label{fig:prompt_optimize}
\end{figure*}

\begin{figure*}
\begin{lstlisting}[language=C++, linewidth=\textwidth]
Your task is to optimize the given program. Generate only a plan to optimize the given program. The rewritten program should be semantically equivalent to the original program. Do not use any loops. Systolic array size is 4x4 (DIM=4) and each element is 4bytes.
// Instructions:
1. The exeuction order of the blocks can be changed. Generate the block ordering as a plan. Do not optimize the instructions within block. Return the plan as a list of blocks. 

<insert ISA prompt here>

<insert unoptimized code here>

\end{lstlisting}
\caption{Task description for reordering the blocks of unoptimized code for generating optimized code.}
\label{fig:prompt_reorder}
\end{figure*}

\begin{figure*}
\begin{lstlisting}[language=C++, linewidth=\textwidth]
Below we describe the functions present in the input code.

```
void tiled_matmul_outer_eigen (
    float *A,
    float *B,
    float *C,
    int i, int k, int j,
    bool transpose_A, bool transpose_B
) {
    for (int i_ctr = 0; i_ctr < i; i_ctr++) {
        for (int j_ctr = 0; j_ctr < j; j_ctr++) {
            for (int k_ctr = 0; k_ctr < k; k_ctr++) {
                float A_elem = A_transpose ? A[k][i] : A[i][k];
                float B_elem = B_transpose ? B[j][k] : B[k][j];
                C[i][j] += A_elem * B_elem;
            }
        }
    }
}

void tiled_matmul_outer_eigen_bias (
    float *A,
    float *B,
    float *D,
    float *C,
    int i, int k, int j,
    bool transpose_A, bool transpose_B, bool sub
) {
    for (int i_ctr = 0; i_ctr < i; i_ctr++) {
        for (int j_ctr = 0; j_ctr < j; j_ctr++) {
            if (sub) {
                C[i_ctr][j_ctr] -= D[i_ctr][j_ctr];
            } else {
                C[i_ctr][j_ctr] += D[i_ctr][j_ctr];
            }
            for (int k_ctr = 0; k_ctr < k; k_ctr++) {
                float A_elem = A_transpose ? A[k_ctr][i_ctr] : A[i_ctr][k_ctr];
                float B_elem = B_transpose ? B[j_ctr][k_ctr] : B[k_ctr][j_ctr];
                C[i_ctr][j_ctr] += A_elem * B_elem;
            }
        }
    }
}
```
\end{lstlisting}
\caption{Code implementation of input functions, as described in Section~\ref{sec:matvec}.}
\label{fig:prompt_matvec_code}
\end{figure*}

\begin{figure*}
\begin{lstlisting}[language=C++, linewidth=\textwidth]
Below we describe the functions present in the input code.

```
void tiled_matmul_outer_eigen (
    const Matrix<float, Dynamic, Dynamic, RowMajor>&A,
    const Matrix<float, Dynamic, Dynamic, RowMajor>&B,
    Matrix<float, Dynamic, Dynamic, RowMajor>&C,
    int i, int k, int j,
    bool transpose_A, bool transpose_B) 
'''
tiled_matmul_outer_eigen performs a matrix multiplication between a matrix in DRAM, represented as A and a matrix in DRAM, represented as B. The result is stored in DRAM, represented as C. 
The dimensions of A are i by k, the dimensions of B are k by j, and the dimensions of C are i by j. transpose_A and transpose_B are boolean values that represent whether the matrix A and B are transposed respectively.
Matrix size is represented as rows x cols, but matrices may be transposed.
'''

void tiled_matmul_outer_eigen_bias (
    const Matrix<float, Dynamic, Dynamic, RowMajor>&A,
    const Matrix<float, Dynamic, Dynamic, RowMajor>&B,
    Matrix<float, Dynamic, Dynamic, RowMajor>&D,
    Matrix<float, Dynamic, Dynamic, RowMajor>&C,
    int i, int k, int j,
    bool transpose_A, bool transpose_B, bool sub)
'''
tiled_matmul_outer_eigen_bias performs a matrix multiplication between a matrix in DRAM, represented as A, and a matrix in DRAM, represented as B. It also adds a bias, stored in DRAM and represented as D, to the final output.
The bias is added if sub is false, and subtracted if sub is true. The result is stored in DRAM, represented as C. 
The dimensions of A are i by k, the dimensions of B are k by j, the dimensions of D are i by j, and the dimensions of C are i by j. transpose_A and transpose_B are boolean values that represent whether the matrix A and B are transposed respectively.
Matrix size is represented as rows x cols, but matrices may be transposed.
'''
```
\end{lstlisting}
\caption{Natural language descriptions of input functions, as described in Section~\ref{sec:matvec}.}
\label{fig:prompt_matvec_code}
\end{figure*}

\begin{figure*}
\begin{lstlisting}[language=C++, linewidth=\textwidth]
Example 1:
#test function
// Multiplication of 4x12 matrix Bdyn, transposed, and 12x1 vector p, not transposed. The matrix and vector are both stored in dram. The result is stored in the 4x1 vector B_p. Systolic array size is 4x4 and each element is 4bytes.
void test(Bdyn, p, B_p) {
    tiled_matmul_outer_eigen(Bdyn, p, B_p, 4, 12, 1, true, false);
}

// rewritten program
```
void test(Bdyn, p, B_p) {
    config_ex(WEIGHT_STATIONARY,  NO_ACTIVATION, true, false);
    config_st(1 * sizeof(float)); // output B_p has 1 column in DRAM
    config_ld(4 * sizeof(float), 0); // A matrix Bdyn has 4 columns in DRAM, because it is transposed
    config_ld(1 * sizeof(float), 1); // B matrix p has 1 column in DRAM
    // Bdyn_sp_addr is the address of the scratchpad where the matrix Bdyn is stored
    static uint32_t Bdyn_sp_addr = 0; // 12 rows, 0 to 11
    // p_sp_addr is the address of the scratchpad where the vector p is stored
    static uint32_t p_sp_addr = 12; // 12 rows, 12 to 23
    // B_p_acc_addr is the address of the accumulator where the output B_p is stored
    static uint32_t B_p_acc_addr = 1 << 31; // 4 rows, 0 to 3
    mvin(Bdyn, Bdyn_sp_addr, 12, 4); // mvin Bdyn as A matrix, 4 rows, 12 cols
    mvin2(p + 0x0, p_sp_addr, 1, 4); // mvin the first 4x1 block of p, 4 rows, 1 cols
    preload(p_sp_addr, B_p_acc_addr, 1, 4, 1, 4); // preload p as matrix B
    compute_preloaded(Bdyn_sp_addr, 0xffffffff, 4, 4, 1, 4); // multiply the first 4x4 block of Bdyn with the first 4x1 block of p
    mvin2(p + 0x4, p_sp_addr + 4, 1, 4); // mvin the second 4x1 block of p, 4 rows, 1 cols
    preload(p_sp_addr + 4, B_p_acc_addr | 1 << 30, 1, 4, 1, 4); // | 1 << 30 since we are accumulating on the same block of B_p
    compute_preloaded(Bdyn_sp_addr + 4, 0xffffffff, 4, 4, 1, 4); // multiply the second 4x4 block of Bdyn with the second 4x1 block of p
    mvin2(p + 0x8, p_sp_addr + 8, 1, 4); // mvin the third 4x1 block of p, 4 rows, 1 cols
    preload(p_sp_addr + 8, B_p_acc_addr | 1 << 30, 1, 4, 1, 4); // | 1 << 30 since we are accumulating on the same block of B_p
    compute_preloaded(Bdyn_sp_addr + 8, 0xffffffff, 4, 4, 1, 4); // multiply the third 4x4 block of Bdyn with the third 4x1 block of p
    mvout(B_p, B_p_acc_addr, 1, 4); // 4 rows is less than or equal to DIM, 1 col, so we can mvout with one instruction
    fence();
}
```
\end{lstlisting}
\caption{The example matrix-vector multiplication code provided for in-context learning, as described in Section~\ref{sec:experiments_gemmini}.}
\end{figure*}
\begin{figure*}
\begin{lstlisting}[language=C++]
Example 2:
#test function
// Multiplication of 12x4 matrix BPA, transposed, and 4x12 matrix Kt, not transposed, minus 12x12 bias matrix Q. The matrices are all stored in DRAM. The result is stored in the 12x12 matrix APBK_Q.
void test(BPA, Kt, Q, APBK_Q) {
    tiled_matmul_outer_eigen_bias(BPA, Kt, Q, APBK_Q, 12, 4, 12, true, false, true);
}
\end{lstlisting}
\caption{Function-specific prompt for a matrix-matrix multiplication from Section~\ref{sec:matmul}.}
\label{fig:prompt_matmat_prob}
\end{figure*}

\begin{figure*}
\begin{lstlisting}[language=C++]
// Configuration for the systolic array
config_ex(WEIGHT_STATIONARY, NO_ACTIVATION, true, false);

// Configure scratchpad and accumulator addresses
static uint32_t BPA_sp_addr = 0;
static uint32_t Kt_sp_addr = 12 * 4; // offset by 12 rows (each row 4 elements wide)
static uint32_t APBK_Q_acc_addr = 1 << 31;
static uint32_t Q_sp_addr = 24 * 4; // offset by 24 rows (each row 4 elements wide)

// Configure loading and output dimensions
config_ld(4 * sizeof(float), 0); // BPA has 4 columns in DRAM when transposed
config_ld(12 * sizeof(float), 1); // Kt has 12 columns in DRAM
config_st(12 * sizeof(float)); // APBK_Q has 12 columns in DRAM

// Move BPA, Kt, and Q into the scratchpad
mvin(BPA, BPA_sp_addr, 4, 12);
mvin2(Kt, Kt_sp_addr, 12, 4);
mvin3(Q, Q_sp_addr, 12, 12);

// Perform tiled matrix multiplication and bias subtraction
for (int i = 0; i < 12; i += 4) { // Iterate over rows of BPA and APBK_Q in blocks of 4
    for (int j = 0; j < 12; j += 4) { // Iterate over columns of Kt and APBK_Q in blocks of 4
        for (int k = 0; k < 4; k += 4) { // Iterate over K dimension (internal dimension of multiplication)
            // Preload Kt blocks
            preload(Kt_sp_addr + k * 12 + j, APBK_Q_acc_addr + i * 12 + j, 4, 4, 4, 4);

            // First computation is non-accumulating
            if (k == 0) {
                compute_preloaded(BPA_sp_addr + i, Q_sp_addr + i * 12 + j, 4, 4, 12, 4);
            } else {
                compute_accumulated(BPA_sp_addr + i, Q_sp_addr + i * 12 + j, 4, 4, 12, 4);
            }
        }
        // After completing K dimension, output the result to DRAM
        mvout(APBK_Q + i * 12 + j, APBK_Q_acc_addr + i * 12 + j, 12, 4);
    }
}

// Ensure all memory operations are completed before proceeding
fence();
\end{lstlisting}
\caption{An example of correct generated code for the function described in Figure~\ref{fig:prompt_matmat_prob}.}
\label{fig:prompt_matmat_sol}
\end{figure*}

\begin{figure*}
\begin{lstlisting}
# Prompt 1
"The solution is slightly incorrect. Identify a few constants that could be added or replaced to make the code correct, for example multipliers for iterators or sizes of operations. Replace the code with <CONST> or define your own constants."

# Prompt 2
"Fill in the constants with the correct values. Options can be selected from the set {0, 1, 3, 4, 12}."
\end{lstlisting}

\caption{The specific prompts used for accelerator-targeted multi-step code repair in Section~\ref{sec:repair}.}
\label{fig:correction_prompt}
\end{figure*}

\clearpage
\section{LLM Autoscheduling}
\label{sec:llm_autoscheduling}

\begin{figure*}[h]
\begin{lstlisting}[linewidth=\textwidth]
SYSTEM: You are an expert performance engineer with experience in optimizing numerical linear algebra kernels.

USER: 
I need help with optimizing a numerical kernel. It is written in a Python DSL
for code optimization called Exo, which is similar to Halide.

Here are my relevant hardware details:

- The target hardware is an x86 CPU with AVX2 support.
- We will be targeting single-core execution, so you can ignore parallelism.
- L1 instruction cache size: 32 KB
- L1 data cache size: 48 KB
- L2 cache size: 2 MB
- L3 cache size: 36 MB

Here is the kernel I need help with, written in Exo:

def doitgen(A: f32[64, 64, 64] @ DRAM, C4: f32[64, 64] @ DRAM,
            sum: f32[64] @ DRAM):
    for r in seq(0, 64):
        for q in seq(0, 64):
            for p in seq(0, 64):
                sum[p] = 0.0
                for s in seq(0, 64):
                    sum[p] += A[r, q, s] * C4[s, p]
            for p in seq(0, 64):
                A[r, q, p] = sum[p]

Currently I get 5.64 GFLOPS.

Please provide a step-by-step plan for optimizing the kernel.
Once you have a plan, begin optimizing the kernel by giving me a series of
commands, each of which are described below. I will apply the command one at a
time, and provide you with the new kernel code and its performance.

You can use the following commands:
```json
{"optimization": "tile", "description": "tile the loop at `line` with `tile_size`", "arguments": {"line": "source code of line to apply tiling", "tile_size": "size of the tile", "outer_name": "name of the new outer loop variable", "inner_name": "name of the new inner loop variable"}}
{"optimization": "fuse", "description": "fuse the loops at `line1` and `line2`", "arguments": {"line1": "source code of first line to fuse", "line2": "source code of second line to fuse"}}
{"optimization": "reorder", "description": "reorder the outer loop at `line` to be the inner loop", "arguments": {"line": "source code of outer loop to reorder"}}
{"optimization": "fission", "description": "fission the loop containing `line` into two consecutive loops, splitting at `line`", "arguments": {"line": "source code of line to fission at", "location": "split (before | after) `line`"}}
{"optimization": "unroll", "description": "fully unroll the loop at `line`", "arguments": {"line": "source code of loop to unroll"}}
```

When you are ready to give the optimization, end your message with JSON as follows:
```
APPLY: {"optimization": "optimization name", "arguments": {"arg1": "value1", ...}}
```
Only give a single command at a time. Do not give multiple commands in a single
message.

Important Notes:
- Consider reordering loops before applying tiling or vectorization.
- Reorder loops to maximize data reuse and minimize cache misses.
- Assume matrices are stored in row-major order.
- If the same source line appears multiple times in the code, end the line
  with ` #N` where `N` is the occurrence number (0-indexed). For example,
  `for i in seq(0, 32): #0`.
\end{lstlisting}
\caption{Task description for scheduling the \texttt{doitgen} kernel.}
\label{fig:llm_autoscheduling_prompt}
\end{figure*}

\begin{figure*}[h]
\begin{lstlisting}[linewidth=\textwidth]
ASSISTANT: To optimize the given kernel, we should focus on improving data locality and reducing cache misses. Here's a step-by-step plan:

1. **Loop Reordering**: We'll start by reordering the loops to maximize data reuse. The innermost loop should iterate over the dimension that allows us to access continuous memory locations, which is beneficial for a row-major storage format.

2. **Loop Tiling**: This technique will help in fitting the data accessed by the loops into the L1 or L2 cache, reducing the number of cache misses.

3. **Loop Unrolling**: This can help to increase the instruction level parallelism by executing multiple iterations of a loop simultaneously. It can also reduce the loop overhead.

4. **Loop Fusion**: If applicable, we can combine loops that iterate over the same range to reduce loop overhead and improve cache locality.

5. **Vectorization**: Since the target hardware supports AVX2, we should ensure that the innermost loops are conducive to vectorization by the compiler, which can significantly speed up the computations.

Starting with loop reordering, we aim to access `A` and `C4` in a more cache-friendly manner. Currently, the access pattern to `A` and `C4` in the innermost loop (`sum[p] += A[r, q, s] * C4[s, p]`) suggests that reordering the loops to bring `s` loop outside could be beneficial. This would align the access to `C4` continuously in memory.

Let's start by reordering the loops to bring the `s` loop outside of the `p` loop:

APPLY: {"optimization": "reorder", "arguments": {"line": "for p in seq(0, 64): #1"}}
====================================================================================================

USER: 
An error occurred while applying the optimization:
argument 1, 'nested_loops' to reorder_loops: expected the body of the outer loop to be a single loop, but it was a def doitgen(A: f32[64, 64, 64] @ DRAM, C4: f32[64, 64] @ DRAM,
            sum: f32[64] @ DRAM):
    for r in seq(0, 64):
        for q in seq(0, 64):
            ...
            for p in seq(0, 64):
                A[r, q, p] = sum[p]  # <-- NODE
Please fix the error and try again.

[a few more errors follow...]

ASSISTANT: It appears that my previous attempts to specify the fission location were incorrect. Let's try a different approach to optimize the kernel without relying on fission at this point.

Given the structure of the loops and the constraints we're facing with the fission command, let's attempt to optimize the kernel by tiling the loops instead. Tiling can help improve cache locality and potentially expose more opportunities for optimization.

We'll start by tiling the `p` and `s` loops, which are involved in the computation `sum[p] += A[r, q, s] * C4[s, p]`. This should help keep the accessed data in cache, reducing cache misses. We'll choose a tile size that is likely to fit well within the L1 cache.

Let's tile the `p` loop with a size of 16, which is a common choice for small tiles that fit well in cache lines:

APPLY: {"optimization": "tile", "arguments": {"line": "for p in seq(0, 64): #1", "tile_size": "16", "outer_name": "p_outer", "inner_name": "p_inner"}}
====================================================================================================

USER: 
I have applied the optimization. The new kernel code is as follows:

def doitgen(A: f32[64, 64, 64] @ DRAM, C4: f32[64, 64] @ DRAM,
            sum: f32[64] @ DRAM):
    for r in seq(0, 64):
        for q in seq(0, 64):
            for p in seq(0, 64):
                sum[p] = 0.0
                for s in seq(0, 64):
                    sum[p] += A[r, q, s] * C4[s, p]
            for p_outer in seq(0, 4):
                for p_inner in seq(0, 16):
                    A[r, q,
                      p_inner + 16 * p_outer] = sum[p_inner + 16 * p_outer]

The new code achieves 6.06 GFLOPS. Please give me another optimization to
apply, using the same format as before.
\end{lstlisting}
\caption{LLM autoscheduling excerpt for \texttt{doitgen}, showing erroneous schedule commands and eventual recovery.}
\label{fig:llm_autoscheduling}
\end{figure*}

\begin{figure*}[h]
\begin{lstlisting}[linewidth=\textwidth]
ASSISTANT: It seems that directly fusing the `r` and `q` loops is not feasible due to their structural positions or other constraints in the code. Let's consider another optimization strategy that might be more straightforward to apply given the current loop structure.

Since we've seen some benefits from tiling and unrolling, let's try to further optimize by unrolling another loop. The `s_outer` loop, which controls the outer part of the tiled `s` loop, is a candidate for unrolling. This could potentially reduce loop overhead and increase the efficiency of the computations inside.

Here's the command to fully unroll the `s_outer` loop:

APPLY: {"optimization": "unroll", "arguments": {"line": "for s_outer in seq(0, 4):"}}
====================================================================================================

USER: 
I have applied the optimization. The new kernel code is as follows:

def doitgen(A: f32[64, 64, 64] @ DRAM, C4: f32[64, 64] @ DRAM,
            sum: f32[64] @ DRAM):
    for r in seq(0, 64):
        for q in seq(0, 64):
            for p in seq(0, 64):
                sum[p] = 0.0
                sum[p] += A[r, q, 0] * C4[0, p]
                sum[p] += A[r, q, 1] * C4[1, p]
                ...
                sum[p] += A[r, q, 62] * C4[62, p]
                sum[p] += A[r, q, 63] * C4[63, p]
            for p_outer in seq(0, 4):
                A[r, q, 16 * p_outer] = sum[16 * p_outer]
                A[r, q, 1 + 16 * p_outer] = sum[1 + 16 * p_outer]
                ...
                A[r, q, 14 + 16 * p_outer] = sum[14 + 16 * p_outer]
                A[r, q, 15 + 16 * p_outer] = sum[15 + 16 * p_outer]

The new code achieves 43.33 GFLOPS. Please give me another optimization to
apply, using the same format as before.
\end{lstlisting}
\caption{The LLM finds an effective transformation for \texttt{doitgen}.}
\label{fig:llm_autoscheduling_good}
\end{figure*}

\end{document}